\documentclass[prd,aps,a4paper,superscriptaddress,onecolumn,nofootinbib]{revtex4}
\usepackage{graphicx}
\usepackage{color}
\usepackage{dcolumn}
\usepackage{bm}
\usepackage{slashed}
\usepackage{amsmath}
\usepackage{latexsym}
\usepackage{amssymb}
\usepackage{mathrsfs}
\usepackage{amsfonts}
\usepackage{url}
\usepackage[normalem]{ulem}
\allowdisplaybreaks

\usepackage{comment}

\begin{document}
\title{Probe the gravitational constant variation via the propagation of gravitational waves}

\author{Bing Sun} 
\affiliation{Department of Basic Courses, Beijing University of Agriculture, Beijing 102206, China}
\affiliation{CAS Key Laboratory of Theoretical Physics, Institute of Theoretical Physics,
Chinese Academy of Sciences, Beijing 100190, China}

\author{Jiachen An} 
\affiliation{Institute for Frontiers in Astronomy and Astrophysics, Beijing Normal University, Beijing 102206, China}
\affiliation{Department of Astronomy, Beijing Normal University, Beijing 100875, China}

\author{Zhoujian Cao
\footnote{corresponding author}} \email[Zhoujian Cao: ]{zjcao@amt.ac.cn}
\affiliation{Institute for Frontiers in Astronomy and Astrophysics, Beijing Normal University, Beijing 102206, China}
\affiliation{Department of Astronomy, Beijing Normal University, Beijing 100875, China}
\affiliation{School of Fundamental Physics and Mathematical Sciences, Hangzhou Institute for Advanced Study, UCAS, Hangzhou 310024, China}

\begin{abstract}
The gravitational constant variation means the breakdown of the strong equivalence principle.
As the cornerstone of general relativity, the validity of general relativity can be examined by studying the gravitational constant variation.
Such variations have the potential to affect both the generation and propagation of gravitational waves.
In this paper, our focus lies on the effect of gravitational constant variation specifically on the propagation of gravitational waves.
We employ two analytical methods, namely based on the Fierz-Pauli action and the perturbation of Einstein-Hilbert action around Minkowski spacetime, both leading to the the same gravitational wave equation.
By solving this equation, we find the effects of gravitational constant variation on gravitational wave propagation.
The result is consistent with previous investigations based on Maxwell-like equations for gravitational waves.
Notably, we find that small variations in the gravitational constant result in an amplitude correction at the leading order and a phase correction at the sub-leading order for gravitational waves.
These results provide valuable insights for probing gravitational constant variation and can be directly applied to gravitational wave data analysis.
\end{abstract}

\maketitle

\tableofcontents

\section{Introduction}
\label{sec:intro}
Among the four fundamental interactions, gravity stands as the first one discovery by humanity due to its prominent presence in early observations and experiments.
Throughout the past century, significant progress has been made in our understanding of the other three fundamental interactions, while gravity has remained relatively enigmatic.
This discrepancy arising from the most advanced facility for experiments, accelerators, is hard to touch gravity.
However, the situation is now changed for gravity, as the advent of gravitational waves has ushered in a new era of exploration.
Gravitational wave astronomy provides us with invaluable detail data pertaining to gravity's behavior in extreme and dynamic environments.
With the progressive advancement of gravitational wave astronomy, these observations amass an ever-growing wealth of information.
To illustrate the remarkable potential of gravitational wave observations in unraveling the gravitational interaction, we can point to the compelling example of GW170817, which effectively rules out numerous alternative gravity theories~\cite{PhysRevLett.119.251301,PhysRevLett.119.251302,PhysRevLett.119.251303,PhysRevLett.119.251304}.

Many different alternative gravity theories predict the variation of effective gravitational constant \cite{2002ApJ...570..463L,PhysRevD.97.104068,2003sttg.book.....F}.
The constraint of gravitational constant variation can be used to distinguish different gravity theories.
In practice, instead of investigating different theories one by one, it is more convenient to study the effect of the gravitational constant variation on gravitational waves in general.

The effect of gravitational constant variation has a notable impact on both the generation and propagation stages of gravitational waves.
Regarding the generation stage, the post-Newtonian approximation provides valuable insights into how gravitational constant variation influences gravitational waves \cite{WANG2022137416,PhysRevD.107.064073}. However, as delve into the propagation stage, the problem is significantly more complex. It is fine to construct the gravitational wave equation based on a specific gravity theory, say general relativity and any other alternative theories.
In the realm of general relativity, the gravitational constant plays a role solely through the interaction between the metric and matter, while it holds no significance for source-free gravitational waves.
However, distinguishing the pure effect of gravitational constant variation from other specific properties within alternative theories presents a challenging task.

In a recent work, we assume that the behaviour of gravitational waves can be aptly captured by a Maxwell-like equation in the presence of gravitational constant variation.
Based on the analysis of such a Maxwell-like equation we found that the gravitational constant variation can be constrained strongly by gravitational wave data even just leading order of small variation of the gravitational constant is considered.
Through a meticulous analysis of this equation, we have uncovered a compelling revelation that the constraints imposed by gravitational wave data are remarkably robust even when considering only a small variation in the gravitational constant at the leading order.
The power of this effect arises from its cumulative impact over vast cosmic distance, spanning from the celestial origin to the detector.
It is worth noting, however, that in the realm of far source-free region and limited to the leading order post-Newtonian approximation, both the gravitoelectric and gravitomagnetic fields are pure gauge effect.
Consequently, it becomes extraordinarily captivating to explore the effect of gravitational constant variation during the gravitational wave propagation in a more fundamental and robust manner.

In this paper, we employ the action principle as our fundamental framework to embark on a rigorous exploration of the influence of gravitational constant variation on the propagation of gravitational waves.
By considering the variation of the gravitational constant at the action level, we investigate two individually independent action principles.
One is the Fierz-Pauli action, the detailed analysis of which is expounded upon in the subsequent section. Another analysis is based on the Einstein-Hilbert action for Minkowski perturbation, elaborating on the results in Sec.~\ref{sec3}.
Intriguingly, we find essentially the same gravitational wave propagation equation for these two actions.
We then conduct a meticulous examination of this equation in Sec.~\ref{sec4}, revealing amplitude corrections and phase corrections in comparison to standard general relativity.
Moreover, along our analysis it yields a compelling constraint on the time variation rate of the gravitational constant $G$ based on GW170817 event.
Finally, we give a summary and discussion in the concluding section.

Throughout the paper we use units with the light speed $c=1$.

\section{Fierz-Pauli action analysis}
\label{sec2}
The Fierz-Pauli action is the well known  and well-established theoretical framework to characterize the behaviour of a relativistic and symmetric rank-2 tensor field living in four-dimensional flat space-time~\cite{PhysRevD.82.044020,2021EPJC...81..171G}, as proposed in~\cite{PhysRevD.102.084045},  which is given by
\begin{equation}
\begin{aligned}
S = & \, \int d^4 x\frac{1}{64 \pi G}\left(-\partial_a h_{b c} \partial^a h^{b c}+\partial_a h_b^b \partial^a h_c^c-2 \partial_a h_c^a \partial^c h_b^b+2 \partial_a h_c^a \partial_b h^{b c}\right).\label{origact}
\end{aligned}
\end{equation}
In our analysis, we consider the variation of $G$ with respect to both time and spatial coordinates. Furthermore, we introduce the trace-inverse tensor defined as
\begin{align}
\bar{h}_{ab}=h_{ab}-\frac{1}{2}h\eta_{ab},
\end{align}
where $h$ is the trace of $h_{ab}$.
By imposing the Lorenz gauge condition $\partial^b \bar{h}_{ab} = 0$, the action \eqref{origact} can be simplified to the form
\begin{equation}
\begin{aligned}
S = & \,\int d^4 x\frac{1}{64 \pi G} \left( -\partial_a\bar{h}_{bc}\partial^a\bar{h}^{bc} + \frac{1}{2}\partial_c\bar{h}\partial^c\bar{h} \right).
\end{aligned}
\end{equation}
The equation of motion derived from this action is given by
\begin{align}
\square h_{ab} - \partial_ch_{ab}\partial^c\ln G = 0. \label{fereom}
\end{align}
In the case where the gravitational constant $G$ is constant, the aforementioned equation reduces to the well-known form of general relativity. Therefore, the equation can be regarded as a generalization from general relativity to scenarios involving a varying gravitational constant.
Additionally, it is important to highlight that the equation is in agreement with the form we obtained previously from Maxwell-like equations that describe the propagation of gravitational waves(Eq.~[46] in~\cite{An}).

The behaviour of gravitational wave propagation could be analyzed utilizing Eq.~\eqref{fereom}. We consider a plane gravitational wave travelling along the $z$-axis, implying $h_{ab} = h_{ab}(t,z)$. Substituting this condition into Eq.~\eqref{fereom}, we obtain the following form
\begin{align}
-\partial_t^2 h_{ab}+\partial_z^2 h_{ab} + \partial_t h_{ab} \partial_t\ln G - \partial_z h_{ab} \partial_z\ln G = 0. \label{ferorg}
\end{align}

To begin, we consider the assumption that the gravitational constant solely depends on time. In this situation, we introduce the separation-of-variables form $h_{ab} = K_{ab}(t)e^{ikz}$, resulting in the following equation
\begin{align}
\partial_t^2 K_{ab}+k^2 K_{ab} - \partial_t K_{ab} \partial_t\ln G = 0. \label{eq1}
\end{align}
Next, we consider the assumption that the gravitational constant depends only on spatial coordinates. Under this assumption, we introduce the separation-of-variables form $h_{ab} = H_{ab}(z)e^{-i\omega t}$, yielding the following equation
\begin{align}
\partial_z^2 H_{ab} + \omega^2 H_{ab} - \partial_z H_{ab} \partial_z\ln G = 0. \label{eq2}
\end{align}
It is intriguing to observe that both Eq.~\eqref{eq1} and Eq.~\eqref{eq2} exhibit exactly identical forms.

\section{Einstein-Hilbert action analysis}
\label{sec3}
In this section we consider the perturbation of Einstein-Hilbert action around the Minkowski spacetime
\begin{align}
g_{ab}=\eta_{ab}+h_{ab}.
\end{align}
We expand the Lagrangian density of Einstein-Hilbert action up to second order
\begin{equation}
\begin{aligned}
\mathcal{L}_{\mathrm{GR}} = \sqrt{-g}R = & -\frac{1}{4}\partial_b h \partial^b h - \partial_a h^{ab}\partial_c h^c_b + \partial_b h\partial_c h^{bc} + \frac{1}{2} h\partial_b\partial_c h^{bc} \\
& + h^{ab}(\partial_a\partial_b h-\partial_b\partial_c h^c_a - \partial_c\partial_b h^c_a + \square h_{ab}) - \frac{1}{2} h \square h - \frac{1}{2}\partial_b h_{ac}\partial^c h^{ab} + \frac{3}{4}\partial_c h_{ab}\partial^c h^{ab}.
\end{aligned}
\end{equation}
By utilizing the trace-inverse tensor and imposing the Lorenz gauge, the above equation can be expressed as follows
\begin{equation}
\begin{aligned}
\sqrt{-g}R = & -\frac{1}{4}\partial_b h \partial^b h - \partial_a h^{ab}\partial_c h^c_b + \partial_b h\partial_c h^{bc} + \frac{1}{2}h\partial_b\partial_c h^{bc} \\
& + h^{ab}(\partial_a\partial_b h-\partial_b\partial_c h^c_a - \partial_c\partial_b h^c_a + \square h_{ab}) - \frac{1}{2}h \square h - \frac{1}{2}\partial_b h_{ac}\partial^c h^{ab} + \frac{3}{4}\partial_c h_{ab}\partial^c h^{ab} \\
= & -\frac{1}{4}\partial_bh\partial^b h - \frac{1}{4}\partial^bh\partial_bh + \frac{1}{2}\partial_bh\partial^bh + \frac{1}{4} h\square h \\
& + h^{ab}\partial_a\partial_b h - h^{ab}\partial_b\partial_a h + \frac{1}{2} h^{ab}\square h_{ab} - \frac{1}{2} h\square h - \frac{1}{2} \partial_b h_{ac}\partial^c h^{ab} + \frac{3}{4}\partial_c h_{ab}\partial^c h^{ab}\\
= & -\frac{1}{4} h\square h + \frac{1}{2} h^{ab}\square h_{ab} - \frac{1}{2} \partial_b h_{ac}\partial^c h^{ab} + \frac{3}{4}\partial_c h_{ab}\partial^c h^{ab} \\
= & -\frac{1}{4} \bar{h}\square \bar{h} + \frac{1}{2}\bar{h}^{ab}\square \bar{h}_{ab} - \frac{1}{2}\partial_b \bar{h}_{ac}\partial^c \bar{h}^{ab} - \frac{1}{8}\partial_b\bar{h} \partial^b\bar{h} + \frac{3}{4}\partial_c \bar{h}_{ab}\partial^c \bar{h}^{ab}.
\end{aligned}
\end{equation}
Consequently the action for gravitational waves becomes
\begin{equation}
\begin{aligned}
S = & \, \int d^4x\frac{1}{32\pi G}\left[  -\frac{1}{2} \bar{h}\square \bar{h} + \bar{h}^{ab}\square \bar{h}_{ab} - \partial_b \bar{h}_{ac}\partial^c \bar{h}^{ab} - \frac{1}{4}\partial_b\bar{h} \partial^b\bar{h} + \frac{3}{2}\partial_c \bar{h}_{ab}\partial^c \bar{h}^{ab} \right].
\end{aligned}
\end{equation}
Under TT gauge condition, we obtain the following expression:
\begin{equation}
\begin{aligned}
S = & \int d^4x \frac{1}{32 \pi G}\left[ \bar{h}^{ij}\square \bar{h}_{ij} - \partial_j \bar{h}_{ik}\partial^k \bar{h}^{ij} + \frac{3}{2}\partial_c \bar{h}_{ij}\partial^c \bar{h}^{ij} \right]. \\
\end{aligned}
\end{equation}
As a result, the equation of motion can be written as
\begin{equation}
\begin{aligned}
\square \bar{h}_{ij} + 2G\square \left(\frac{1}{G} \right)\bar{h}_{ij} - \partial^a \ln G \partial_a\bar{h}_{ij} - 4\partial^k \ln G\partial_j \bar{h}_{ik} = 0.
\end{aligned}
\end{equation}

Similarly to Sec.\ref{sec2}, we assume that the gravitational wave propagates as a plane wave along the $z$-axis without loss of generality. This implies that $\bar{h}_{ij}$ depends solely on $t$ and $z$.

Firstly assuming that $G$ only depends on time, the equation of motion takes the following form
\begin{equation}
\begin{aligned}
\square \bar{h}_{ij} - 2G\partial_t^2 \left(\frac{1}{G} \right)\bar{h}_{ij} + \partial_t\ln G \partial_t\bar{h}_{ij} = 0.
\end{aligned}
\end{equation}
Then we introduce variable separation by letting $\bar{h}_{ij} = H_{ij}(t)e^{ikz}$.
This yields the following expression
\begin{equation}\label{eht}
\begin{aligned}
\partial_t^2H_{ij} - \partial_t\ln G\partial_tH_{ij} + \left[k^2 + 2G\partial_t^2 \left(\frac{1}{G} \right)\right] H_{ij} = 0.
\end{aligned}
\end{equation}

Secondly, assuming that $G$ is only spatially dependent, we obtain the following differential equations for $\bar{h}_{ij}$
\begin{equation}
\begin{aligned}
\square \bar{h}_{ij} + 2G\nabla^2 \left(\frac{1}{G}\right) \bar{h}_{ij} - \partial^k \ln G \partial_k \bar{h}_{ij} - 4\partial^k \ln G\partial_j \bar{h}_{ik} = 0.
\end{aligned}
\end{equation}
We then separate the variables as $\bar{h}_{ij} = H_{ij}(z)e^{-i\omega t}$, resulting in the spatial differential equations
\begin{equation}
\begin{aligned}
\nabla^2H_{ij} + \left[ \omega^2 + 2G\nabla^2 \left(\frac{1}{G}\right)\right]H_{ij} - \partial^k \ln G \partial_k H_{ij} - 4\partial^k \ln G\partial_j H_{ik} = 0.
\end{aligned}
\end{equation}
Because the GW travels along $z$-axis, the $H_{ij}$ is the function of $z$ only,
\begin{equation}
\begin{aligned}
    \partial_z^2 H_{ij} + \left[ \omega^2 + 2G\partial_z^2 \left(\frac{1}{G}\right) \right]H_{ij} - \partial_z \ln G \partial_z H_{ij} - 4\delta^z_j\partial^k \ln G\partial_z H_{ik} = 0.
\end{aligned}
\end{equation}
Within TT gauge, $H_{ij}$ furthermore only has $+$ and $\times$ components so the equations become
\begin{equation}\label{ehz}
\begin{aligned}
    \partial_z^2 h- \partial_z \ln G\partial_z h + \left[\omega^2 + 2G\partial_z^2 \left(\frac{1}{G}\right) \right]h  = 0,
\end{aligned}
\end{equation}
where $h$ represents $h_+$ or $h_\times$.
Approximately this equation means that the gravitational wave admits a frequency shift $\frac{G\omega}{\omega^2}\partial_z^2\left(\frac{1}{G}\right)$ compared to the case of a constant $G$.
We note that the above equation exhibits exactly the same form as Eq.~(\ref{eht}).

\section{The gravitational wave equation analysis}\label{sec4}
We have obtained Eqs.~\eqref{eq1}, \eqref{eq2}, \eqref{eht}, and \eqref{ehz} to describe the propagation behavior of gravitational waves.
Remarkably, all four equations can be unified and expressed in the form of
\begin{equation}
    \frac{d^2}{d\zeta^2}H(\zeta)+2p(\zeta)\frac{d}{d\zeta}H(\zeta)+\left[\xi^2+q(\zeta)\right]H(\zeta)=0.\label{dessoleq}
\end{equation}
Here, $\zeta$ is a generic self variable and does not correspond to any specific coordinate.
The parameter $\xi$ represents either the wave number $k$ (in Eqs.~\eqref{eq1} and \eqref{eht}) or the wave frequency $\omega$ (in Eqs.~\eqref{eq2} and \eqref{ehz}).
The unknown function $H$ represents any of the unknown functions shown in Eqs.~\eqref{eq1}, \eqref{eq2}, \eqref{eht}, and \eqref{ehz} respectively.
More concretely, we have the following expressions
\begin{equation}
p=-\frac{1}{2}\partial_t \ln G,\quad q=0,\label{eq3}
\end{equation}
for Eq.\eqref{eq1},
\begin{equation}
p=-\frac{1}{2}\partial_z \ln G,\quad q=0,\label{eq4}
\end{equation}
for Eq.\eqref{eq2},
\begin{equation}
p=-\frac{1}{2}\partial_t \ln G,\quad q=2G\partial_t^2\left(\frac{1}{G}\right),\label{eq6}
\end{equation}
for Eq.\eqref{eht}, and
\begin{equation}
    p=-\frac{1}{2}\partial_z \ln G,\quad q=2G\partial_z^2 \left(\frac{1}{G}\right),\label{eq7}
\end{equation}
for Eq.\eqref{ehz}.

So in order to analyze the effect of gravitational constant variation on the propagation of gravitational waves, we need only investigate Eq.\eqref{dessoleq}.

We can use two real functions $A(\zeta)$ and $\Phi(\zeta)$ to decompose the complex function $H(\zeta)$ as
\begin{equation}\label{HAphi}
    H=Ae^{i\Phi}.
\end{equation}
We then plug Eq.~\eqref{HAphi} into Eq.~\eqref{dessoleq} and divide both sides of the equation by $e^{i\Phi}$.
From the real and imaginary parts of the result, we achieve two equations,
\begin{equation}\label{rpart}
    \frac{d^2 A}{d\zeta^2}+2p\frac{d A}{d\zeta}+\left[\xi^2\left(1-\frac{B^2}{\xi^2}\right)+q\right]A=0,
\end{equation}
\begin{equation}\label{ipart}
    2\frac{d A}{d\zeta}B+A\frac{d B}{d\zeta}+2pAB=0.
\end{equation}
Here $B\equiv\frac{d \Phi}{d\zeta}$.
The two unknown functions in the above differential equations are denoted as $A$ and $B$. It is evident that if $(A, B)$ is a solution to the linear equations, $(A, -B)$ is another linearly independent solution. Therefore, we focus on the positive $B$ solution without loss of generality.

From \eqref{ipart} we have
\begin{align}
&    2\frac{1}{A}\frac{d A}{d\zeta}+\frac{1}{B}\frac{d B}{d\zeta}+2p=0,\\
&    2\frac{d \ln A}{d\zeta}+\frac{d \ln B}{d\zeta}+2p=0,\\
&    \frac{d \ln A^2+\ln B}{d\zeta}=-2p,\\
&    \ln(A^2B)=C-2\int pd\zeta.
\end{align}
Here $C$ is an integral constant. Therefore we have
\begin{equation}\label{Apk}
    A=Ce^{-\int p\,d\zeta}B^{-1/2}.
\end{equation}
We plug Eq.~\eqref{Apk} into Eq.~\eqref{rpart} and then achieve
\begin{align}
    &\frac{d^2 K}{d\zeta^2}-\left(\frac{1}{\Gamma}\frac{d^2\Gamma}{d\zeta^2}-q\right)K+\xi^2K(1-K^{-4})=0,\label{equK0}\\
    &\Gamma\equiv e^{\int p \,d\zeta},\label{eq5}
\end{align}
where $K\equiv(B/\xi)^{-1/2}$.

When we have already got the solution $K$ of the above Eq.~(\ref{equK0}) we can easily construct the two linear independent solutions of $B$ as
\begin{align}
B=\pm\xi K^{-2}
\end{align}
according to the discussion after Eq.~(\ref{ipart}). Based on these two linear independent solutions we can express the general solution to Eq.~\eqref{dessoleq} as
\begin{equation}
    H=K\Gamma^{-1}\left(C_+e^{i\xi\int  K^{-2}\,d\zeta}+C_-e^{-i\xi\int  K^{-2}\,d\zeta}\right),
\end{equation}
where $K$ is the solution of Eq.~\eqref{equK0}, and $C_{\pm}$ are two integral constants. Note that the two components correspond to waves propagating along and anti-along $z$ direction. Since the real gravitational waves always propagating from source to detector, we need only leave one component. Without losing any generality, we can express the solution corresponding to real gravitational waves as
\begin{equation}
    H=C_+K\Gamma^{-1}e^{i\xi\int  K^{-2}\,d\zeta}.
\end{equation}

Now we pay particular attention to the behavior of $K$. We introduce notation
\begin{align}
\Xi\equiv\frac{1}{\Gamma}\frac{d^2\Gamma}{d\zeta^2}-q.
\end{align}
And more we use unit to make $\xi=1$, then Eq.~(\ref{equK0}) can be simplified as
\begin{equation}\label{equK}
    \frac{d^2 K}{d\zeta^2}+K[(1-\Xi)-K^{-4}]=0.
\end{equation}
If $\Xi$ in the above equations is a constant, it is easy to find a particular solution
\begin{equation}\label{K0}
    K=(1-\Xi)^{-1/4}=1+\frac{1}{4}\Xi+\frac{5}{32}\Xi^2+O(\Xi^3).
\end{equation}

Considering that $G$ must be slowly varying at most, we can reform $\Xi$ as
\begin{align}
&\Xi(\zeta)=\kappa^2\tilde{\Xi}(\tilde{\zeta}),\\
&\tilde{\zeta}=\kappa\zeta
\end{align}
where $\kappa$ is a small parameter. Then Eq.~(\ref{equK}) becomes
\begin{equation}\label{equKs}
    K^3\frac{d^2 K}{d \tilde{\zeta}^2}\kappa^2-K^4\tilde{\Xi}(\tilde{\zeta})\kappa^2+K^4-1=0.
\end{equation}
We suppose that
\begin{equation}\label{K}
    K=\sum_{n=0}^\infty K_n(\tilde{\zeta})\kappa^{2n}.
\end{equation}
We then plug Eq.~\eqref{K} into Eq.~\eqref{equKs}, and expand the left hand side of the equation as a power series of $\kappa^2$. Since each coefficient of this power series should vanish, we have
\begin{gather}
    K_0^4-1=0,\\
    K_0^3K_0''-K_0^4\tilde{\Xi}+4K_0^3K_1=0,\\
    (K_0^3K_1''+3K_0^2K_1K_0'')-4K_0^3K_1\tilde{\Xi}+(4K_0^3K_2+6K_0^2K_1^2)=0,
\end{gather}
for the leading three ones. The prime notation means the derivative respect to $\tilde{\zeta}$.

Finally, by solving all of these equations in order, we will find each of the coefficients in Eq.~\eqref{K}, for example,
\begin{gather}
    K_0=1,\\
    K_1=\frac{1}{4}\tilde{\Xi},\\
    K_2=\frac{5}{32}\tilde{\Xi}^2-\frac{1}{16}\frac{d^2\tilde{\Xi}}{d\tilde{\zeta}^2}.
\end{gather}
These coefficients can be used to construct the particular solution of $K$ according to Eq.~(\ref{K}) and let $\kappa=1$ formally. We can note that when $\Xi$ is a constant, this formal series solution recovers (\ref{K0}) exactly which indicates the reliability of our method.

For convenience of later application, we now recover to usual units without restriction of $\xi=1$. Note that the dimension of $\Xi$ is $\frac{1}{L^2}$, the dimension of $\xi$ is $\frac{1}{L}$ and $K$ is dimensionless, we have
\begin{gather}
    K=1+\frac{1}{4\xi^2}\Xi+\frac{1}{16\xi^4}\left(\frac{5}{2}\Xi^2-\frac{d^2\Xi}{d\zeta^2}\right)+O\left[(\frac{\Xi}{\xi^2})^3\right].
\end{gather}
\section{The effect of the gravitational constant variation on the propagation of gravitational waves}

Now we apply the solution got in the last section to the gravitational wave equations (\ref{eq1}), (\ref{eq2}), (\ref{eht}) and (\ref{ehz}). For all of these four equations we are interested in, we have
\begin{align}
&\Gamma=G^{-1/2},
\end{align}
according to Eq.~(\ref{eq3})-(\ref{eq7}). If we consider the solution $K$ only to leading order respect to $\kappa$ shown in (\ref{K}), we have
\begin{align}
&K=1,\\
&H=\frac{\Gamma_s}{\Gamma_d}H^0=\sqrt{\frac{G_d}{G_s}}H^0,
\end{align}
where $H^0$ means the solution when $G$ does not vary and the subscript $s,d$ mean gravitational wave source and detector. This is to say the amplitude of gravitational waves will be adjusted while leave the phase unchanged. This is to say the amplitude of gravitational waves will be adjusted while leave the phase unchanged. This is consistent to the result we got in previous work \cite{An}.

The analysis done in \cite{An} is based on Maxwell-like equations which may break down to describe source free gravitational wave. In addition, the analysis done in \cite{An} is only valid to spacial point dependent $G$. Here our analysis is more solid and the result is valid for both time dependent cases and spacial point dependent cases. Similar to the data analysis of GW170817 done in \cite{An}, we can extend the result to time variation rate of $G$ here. Based on the posterior distribution of $\frac{G_s}{G_d}$ we got in \cite{An}, we can constraint the time variation rate of $G$ as following.

\begin{figure}
\begin{tabular}{c}
\includegraphics[width=0.45\textwidth]{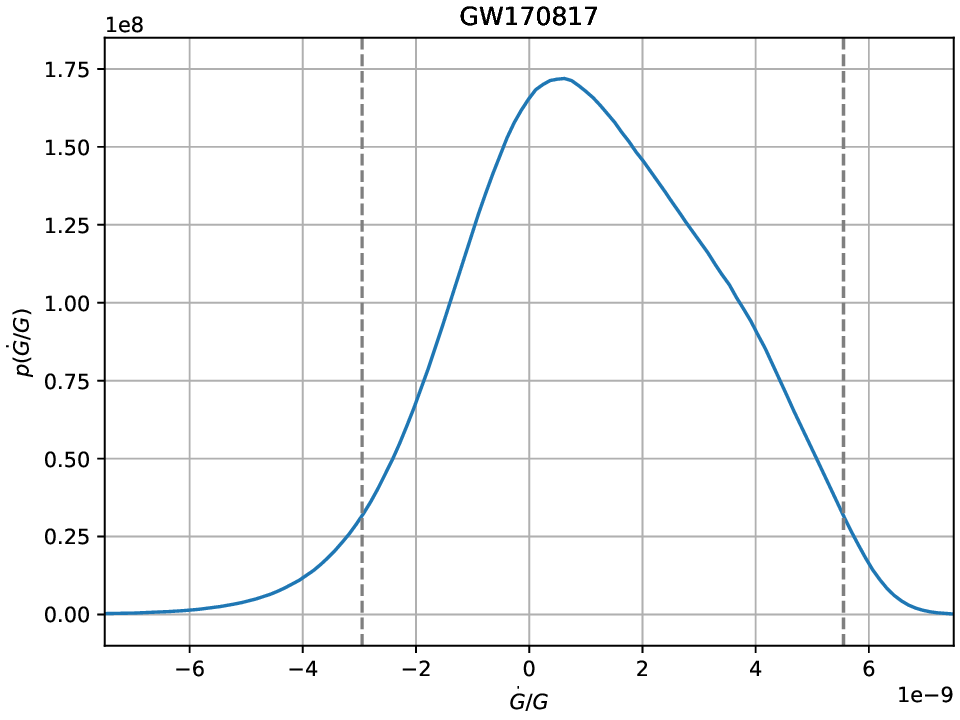}
\end{tabular}
\caption{Posterior distribution of $\frac{\dot{G}}{G}$ according to GW170817/GRB170817A. The region between the two vertical dashed lines represents the parameter domain with 95\% confidence.}\label{fig2}
\end{figure}

If we assume $G$ is slowly varying respect to time, we can estimate
\begin{align}
G_s=G_d-\dot{G}T,
\end{align}
here $T$ corresponds to the cosmological time between GW generation and detection. Then we have
\begin{align}
\frac{G_s}{G_d}=1-\frac{\dot{G}}{G_d}T.\label{eq13}
\end{align}
Regarding to GW170817, $T\sim1.4\times10^8$yr. Consequently we can constrain
\begin{align}
\left|\frac{\dot{G}}{G}\right|\lesssim10^{-9}{\rm yr}^{-1}.
\end{align}
Based on the distribution of $\frac{G_s}{G_d}$ we got above and the relation (\ref{eq13}) we can get the posterior distribution of $\frac{\dot{G}}{G}$. We plot this distribution in Fig.~\ref{fig2}. This constraint is a little bit better than the one got through GW speed \cite{PhysRevLett.126.141104,WANG2022137416}.

If we consider the solution $K$ to sub-leading order, we have
\begin{align}
&K=1+\frac{1}{4\xi^2}\Xi,\\
&\xi\int K^{-2}\,d\zeta=\xi\int (1+\frac{1}{4}\frac{\Xi}{\xi^2})^{-2}\,d\zeta\approx\xi\int (1-\frac{1}{2}\frac{\Xi}{\xi^2})\,d\zeta=\xi \zeta- \frac{1}{2\xi}\int\Xi\,d\zeta.
\end{align}
So we get
\begin{align}
&H=\frac{K_d\Gamma_s}{K_s\Gamma_d}e^{-i\frac{1}{2\xi}\int\Xi\,d\zeta}H^0=\frac{1+\frac{1}{4\xi^2}\Xi_d}{1+\frac{1}{4\xi^2}\Xi_s}\sqrt{\frac{G_d}{G_s}}e^{-i\frac{1}{2\xi}\int\Xi\,d\zeta}H^0.
\end{align}
That is to say we have both amplitude correction $\frac{1+\frac{1}{4\xi^2}\Xi_d}{1+\frac{1}{4\xi^2}\Xi_s}\sqrt{\frac{G_d}{G_s}}$ and phase correction $-\frac{1}{2\xi}\int\Xi\,d\zeta$. Since the wave number $k$ equals to wave frequency $\omega$ for gravitational waves in general relativity, we can summarize the above results as follows. The variation of gravitational constant $G$ introduce amplitude correction factor
\begin{align}
\frac{1+\frac{1}{4\omega^2}\Xi_d}{1+\frac{1}{4\omega^2}\Xi_s}\sqrt{\frac{G_d}{G_s}},
\end{align}
and phase correction term
\begin{align}
&-\frac{1}{2\omega}\int_s^d\Xi\,dr,\\
&\Xi=-\frac{13}{4}G^{-2}(\partial_rG)^2
\end{align}
for position dependent $G$ or
\begin{align}
&-\frac{1}{2\omega}\int_s^d\Xi\,dt,\\
&\Xi=-\frac{13}{4}G^{-2}(\partial_tG)^2
\end{align}
for time dependent $G$. The integration above is taken from source to detector. And the above summary has already neglected second order derivative of $G$ respect to space or time due the small variation fact of $G$.

We note that both the amplitude correction and the phase correction are frequency dependent. This means that the waveform will not changed compared to general relativity theory. Based on the waveform template of general relativity theory, we can do the above adjustment to get waveform template considering the variation of gravitational constant $G$. Use this new waveform template we can investigate the variation behavior of gravitational constant $G$ with the gravitational wave signal only. We plan to do such extensive data analysis in the future.

\section{summary and conclusion}
Spacetime singularities indicate a potential breakdown of general relativity under certain conditions. However, determining a theory that would replace general relativity in such scenarios remains a challenging task. Alternatively, we can explore which fundamental principle of general relativity breaks down when the theory is no longer valid. In this work, we focus on the validity of the strong equivalence principle, which is highly related to the variation of the gravitational constant $G$ from a phenomenological perspective~\cite{RevModPhys.75.403,Uzan2011}.

In the current paper, we have employed the action principle to investigate the behavior of gravitational waves in the context of varying gravitational constant $G$. In order to make our result more solid, we have studied two different actions: the Fierz-Pauli action and the Einstein-Hilbert action. As we expected, the resulted gravitational wave propagation equations are essentially the same. Subsequently, we analyzed the solutions of these equations by assuming a small variation in $G$. Our analysis reveals that at the leading order of such variation, only the amplitude of the gravitational wave is corrected. However, when considering the sub-leading order, both the amplitude and phase are corrected, which are also dependent on the gravitational wave frequency. Based on these detailed correction forms, we can construct waveform template in scenarios where the gravitational constant $G$ is varying, if the waveform of general relativity is known.

In previous studies~\cite{WANG2022137416,PhysRevD.107.064073}, the impact of varying gravitational constant $G$ on the generation of gravitational waves has been well studied. In this work, we extend the investigation to include the effect of $G$ variation on the propagation of gravitational waves. By combining these two distinct effects, we can construct comprehensive waveform templates for the situation involving a varying $G$. This investigation provides valuable insights into the behaviour of the variation of gravitational constant $G$ based on gravitational wave detection results, and contributes to a deeper understanding of the fundamental aspects of gravity.

\bibliography{reference}

\end{document}